\documentclass[a4paper,11pt]{article}
\pdfoutput=1 

\usepackage{jheppub} 

\usepackage[utf8]{inputenc}

\usepackage{tikz}

\usepackage{amsmath,amssymb,amsthm}

\usepackage{color}

\usepackage{framed}

\title{Entanglement entropy through conformal interfaces in the 2D Ising model}
\preprint{LMU-ASC 23/15}

\author[a]{E. Brehm}
\author[a]{and I. Brunner}
\affiliation[a]{Arnold Sommerfeld Center, Ludwig Maximilians Universität\\Theresienstraße 37, 80333 München, Germany}

\emailAdd{E.Brehm@physik.uni-muenchen.de}
\emailAdd{Ilka.Brunner@physik.uni-muenchen.de}

\abstract{We consider the entanglement entropy for the 2D Ising model at the conformal fixed point in the presence of interfaces. More precisely, we investigate the situation where the two subsystems are separated by a defect line that preserves conformal invariance. Using the replica trick, we compute the entanglement entropy between the two subsystems. We observe that the entropy, just like in the case without defects,  shows a logarithmic scaling behavior with respect to the size of the system. Here, the prefactor of the logarithm depends on the strength of the defect encoded in the transmission coefficient. We also comment on the supersymmetric case.}

\newcommand{\fref}[1]{figure \ref{#1}}
\newcommand{\bra}[1]{\langle#1\vert}
\newcommand{\ket}[1]{\vert#1\rangle}
\newcommand{\mc}[1]{\mathcal{#1}}
\newcommand{\IBox}[1]{\begin{center}
                       \parbox{0.95\textwidth}{\begin{framed}
                        #1
                       \end{framed}}
                      \end{center}
}
\newcommand{\Z}{\mathbb{Z}}
\newcommand{\Tr}{\text{Tr}}

\begin{document}
\maketitle
\flushbottom

\section{Introduction and summary}

Entanglement entropy in quantum systems has been investigated in recent years in many different fields, ranging from quantum information theory to black hole physics. It encodes the information of the entanglement of a subsystem $A$ with the rest of the system. In 1+1 dimensional systems at the critical point, the vacuum entanglement entropy for a subsystem $A$ which is obtained by geometrically singling out an interval is given by \cite{calabrese_entanglement_2009,holzhey_geometric_1994,vidal_entanglement_2003}
\begin{equation}\label{eq:basicee}
S_A = - \Tr \rho_A \log \rho_A = \frac{c}{3} \log L,
\end{equation}
where $\rho_A$ is the reduced density matrix and $L$ the length of the interval specifying $A$. 

In this paper, we  consider situations where the separation of the full system is not merely geometric. Rather, we investigate the case where there is a physical separation given by a conformal interface (to which we also refer as ``defect''). This is a one-dimensional domain wall, localized in the time direction, separating the two-dimensional space-time into two parts. An interface CFT naturally consists of two sub-systems -- the two theories that are joined by the defect line. 
The domain wall can be fully or partially transmissive, such that the quantum field theories living on the two sides are related nontrivially across the defect line. A useful and interesting probe to the whole system is the quantum correlation, i.e. entanglement, between the two sub-systems. The subsystem $A$ in  (\ref{eq:basicee}) thus consists of the CFT on one side of the interface.

In the present paper, our discussion focusses on the two-dimensional Ising model, where defects have been analysed by integrability \cite{henkel_ising_1989,abraham_transfer_1989} and conformal field theory \cite{oshikawa_defect_1996,oshikawa_boundary_1997} techniques. There are altogether three classes of defects preserving conformal invariance. Two of them have a simple description in terms of a square lattice model.  At the position of the interface, the couplings between the spins are different than in the bulk of the lattice. In formulas, the energy-to-temperature ratio is given by 

\begin{equation}
 \frac{\mc{E}}{T} = -\sum_{i,j}\left(K_1 \sigma_{i,j}\sigma_{i+1,j} + K_2\sigma_{i,j}\sigma_{i,j+1}\right) + (1-b) K_1 \sum_j \sigma_{0,j}\sigma_{1,j} \label{eq:defE/T}
\end{equation}
where $\sigma_{i,j} = \pm1$ are the spin variables, and $\sinh(2K_1)\sinh(2K_2) = 1$ so that the bulk theory is critical. Along the (vertical) interface, couplings are rescaled by the factor $b$ that parametrizes deformations of the interface. In the special case $b=1$ the situation reduces to the case without any defect, on the other hand, for $b= \pm \infty$ or $b=0$, one obtains two isolated subsystems, separated  by a totally reflective defect. One furthermore distinguishes between ferromagnetic interfaces for which the parameter $b$ takes values  $b\in(0,\infty)$ and anti-ferromagnetic interfaces which are parametrized by $b\in(-\infty,0)$.

In a spin chain interpretation, the defect sits on a particular link of the spin chain, and we consider the entanglement entropy of the subsystems located left and right of the defect link. When the system propagates in time, the defect link sweeps out a one-dimensional line in two-dimensional space-time, which is the defect line of the conformal field theory.

One now expects the entanglement entropy between two subsystems to reduce to \eqref{eq:basicee}  in the totally transmissive case at $b=1$ and to vanish in the totally reflective case. For generic $b$, the entanglement entropy will depend on $b$, as this parameter determines the ``strength'' of the defect.  

Indeed, we will show that the entanglement entropy is given by

\begin{equation}\label{eq:result}
 S = \sigma(\mc{T}) \log L + C,
\end{equation}
where $C$ is a constant independent of $L$ and $\mc{T}$ parametrizes the transmission of the defect. For the defects \eqref{eq:defE/T} $\mc{T}$ can be expressed in terms of $b$ and the constant $C$ vanishes. The formula for the case of the remaining class of  defects, which is not described by  (\ref{eq:defE/T}), is similar. This class can be obtained by appyling order-disorder duality to one side of the interface, which does not change the transmissivity and also not the prefactor $\sigma({\mc{T}})$. On the other hand, it does shift the constant $C$.

Entropy formulas of the form (\ref{eq:result}) have appeared in several different contexts in 2D CFT. First of all, the entropy through a conformal interface was considered in the example of the free boson compactified on a circle in \cite{sakai_entanglement_2008} with a  result  precisely of the form (\ref{eq:result}). Indeed, our analysis in sections \ref{sec:howtoEE},  \ref{sec:derPartition} ,\ref{sec:derEE} is similar to their computations.

The entanglement entropy for subsystems separated by a defect in the Ising model as well as other fermionic chains was studied before with different methods. Numerical results were presented in \cite{2009PhRvB..80b4405I}, subsequently an analytical analysis appeared  in  \cite{2010arXiv1005.2144E}. In particular, the form of the entanglement entropy (\ref{eq:result}) was derived using the spectrum of the reduced density matrix in the lattice model. The paper \cite{2010arXiv1005.2144E} initiated a series of following papers addressing related topics, see e.g. \cite{2012JPhA...45j5206C,2012JPhA...45o5301P,2012EL.....9920001E,2013JPhA...46q5001C,2015arXiv150309116E}.

In the present paper, we investigate the defects of the Ising model from the point of view of conformal field theory. Here, one associates one class of defect lines to each primary of the Ising model. Within each class, the elements differ by marginal perturbations and as a result also by their transmission and we verify (\ref{eq:result}) by a conformal field theory computation. The different classes differ physically by properties such as their $g$-factor and RR-charge. From the point of view of the entanglement entropy, this changes the constant contribution in (\ref{eq:result}). The constant shifts are particularly interesting in the supersymmetric case, which we analyze by combining the Ising-model results with those of  \cite{sakai_entanglement_2008}.

Constant shifts in the entanglement entropy were also observed in \cite{Nozaki:2014hna,Nozaki:2014uaa,He:2014mwa}. In those papers, the shift in the entanglement entropy of excited states, rather than the vacuum, was determined. The excited states were obtained by acting with local operators on a CFT vacuum. The physical difference to our situation is that the defects we consider extend in one dimension, hence are not local operators. In the case of the papers \cite{Nozaki:2014hna,Nozaki:2014uaa,He:2014mwa} the logarithmic term remains the same (compared to the situation of the vacuum), whereas the constant term gets shifted. In the case of rational conformal field theories, the constant shifts have an interpretation in terms of quantum dimensions. In \cite{PandoZayas:2014wsa,Das:2015oha} the entanglement entropy was considered for conformal field theories with boundary. These systems are related to ours by the folding trick, where one folds along the defect line to obtain a tensor product theory with a boundary, see figure \ref{fig:foldingtrick}. However, the division in subsystems is different in their case, as they consider the  division of the system into left and rightmovers and compute the left-right entropy.

This paper is organized as follows: In section  \ref{sec:confInterfaces} we review the construction of conformal interfaces for the Ising model. Using the folding trick, conformal interfaces can be mapped to boundary conditions for a free boson on a circle orbifold. Hence, they are given in terms of D0 branes (ferromagnetic and anti-ferromagnetic) and D1 branes (order-disorder) parametrized by their position and Wilson lines, respectively. While this description offers an intuitive interpretation of the possible interfaces, it is less useful for calculations in our context. We hence go back to a formulation in terms of a GSO projected free fermion theory.
The ferromagnetic and anti-ferromagnetic interfaces are then given by interfaces charged under RR-charge whereas the order-disorder interface is a neutral interface.  

In section \ref{sec:howtoEE} we explain the basics of how to compute the entanglement entropy as a derivative of a partition function involving defects. Sections \ref{sec:derPartition} and \ref{sec:derEE} contain the concrete calculations for the Ising model, in particular $\sigma(\mathcal{T})$ and $C$  in equation (\ref{eq:result}) is derived for all classes of interfaces of the Ising model. In section \ref{sec:SUSY} we comment on the supersymmetric model, combining our results with those of \cite{sakai_entanglement_2008}. We show that $\sigma(\mc{T})$ simplifies due to cancellations between oscillator modes. We also express the result in a way that is suggestive for generalizations to higher dimensional tori and possibly other supersymmetric models. Finally, in section \ref{sec:conclusions} we draw some conclusions and point out open problems. 

\section{Conformal interfaces of the Ising model}
\label{sec:confInterfaces}
\subsection{Interfaces and boundary conditions}
 
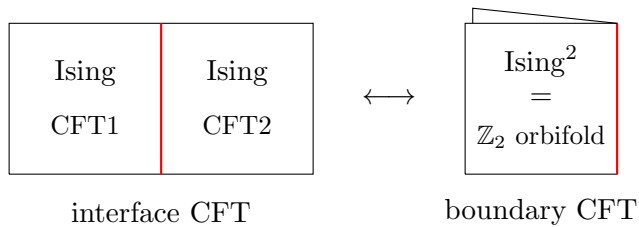
\begin{figure}[ht!]
\begin{center}
  \begin{tikzpicture}[scale=2]
 \coordinate (a) at (0,0);
 \coordinate (b) at (0,1);
 \coordinate (c) at (1,1);
 \coordinate (d) at (1,0);
 \coordinate (2d) at (2,0); 
 \coordinate (2c) at (2,1);
 \node at (.5,.5) {$\begin{matrix}\text{Ising}\\[.5em]\text{\small{CFT1}}\end{matrix}$};
 \node at (1.5,.5) {$\begin{matrix}\text{Ising}\\[.5em]\text{\small{CFT2}}\end{matrix}$};
 \node at (1,-.25) {interface CFT};
 \node at (3.5,-.25) {boundary CFT};
 \node at (2.5,.5) {$\longleftrightarrow$};
 \draw (a) -- (b) -- (2c) -- (2d) -- (a);
 \draw[thick,red] (c) -- (d);
 \coordinate (a) at (3,0);
 \coordinate (b) at (3,1);
 \coordinate (c) at (4,1);
 \coordinate (d) at (4,0);
 \node at (3.5,.75) {$\text{Ising}^2$};
 \node at (3.5,.5) {$=$};
 \node at (3.5,.25) {\small{$\mathbb{Z}_2$ orbifold}};
 \draw (a) -- (b) -- (c) -- (d) -- (a);
 \draw[thick,red] (c) -- (d);
 \draw (3.05,1) -- (3.05,1.1) -- (c);
 \end{tikzpicture}
 \end{center}
 \caption{The folding trick transforms interfaces (red) of the critical Ising model to boundary conditions of the $c=1$ $\Z_2$-orbifold theory.}\label{fig:foldingtrick}
\end{figure}

A convenient description of defects in the Ising model arises, when we employ the folding trick. Here, as illustrated in \fref{fig:foldingtrick}, an Ising model interface is mapped to a boundary condition of the tensor product of two Ising models. The latter is well known to be equivalent to a $\Z_2$ orbifold of a free boson compactified on a circle of radius $r=1$ \cite{oshikawa_defect_1996,oshikawa_boundary_1997}.


The boundary conditions of the orbifold theory come in two continuous families, \cite{oshikawa_defect_1996,oshikawa_boundary_1997,quella_reflection_2007}: 

\begin{itemize}
 \item Dirichlet conditions $\ket{D,\phi}\hspace{-2pt}\rangle$ with $\phi\in [0,\pi]$\,,
 \item Neumann conditions $\ket{N,\tilde \phi}\hspace{-2pt}\rangle$ with $\tilde \phi\in [0,\pi/2]$\,.
\end{itemize}

In string theory language, $\phi$ is the position of a D0-brane on the circle with a $\Z_2$ identification, whereas $\tilde \phi$ is the Wilson line on a D1-brane which belongs to the position of the dual D0-brane on the dual circle (of radius $\tilde r =1/2$). Unfolding converts the boundary states $\ket{\mc{B}}\hspace{-2pt}\rangle$ of (Ising)$^2$ to interfaces of the Ising model. 

For Dirichlet interfaces one can relate $\phi$ and the parameter $b$ of the interface model \eqref{eq:defE/T} as in \cite{oshikawa_defect_1996,oshikawa_boundary_1997} by comparing the CFT spectrum with the exact diagonalization of the transfer matrix \cite{delfino_scattering_1994}:

\begin{equation}
 \tan(\phi-\pi/4) = \frac{\sinh(K_1(1-b))}{\sinh(K_1(1+b))} ~\longleftrightarrow~ \cot(\phi) = \frac{\tanh(bK_1)}{\tanh(K_1)} \label{eq:phi-b}
\end{equation}

A special case is $\phi = \pi/4$ corresponding to $b=1$ which means there is no interface. Hence the interface operator is given by the identity operator. Another special case is $\phi = 3\pi/4$ which belongs to $b=-1$. This operator belongs to the $\Z_2$-symmetry of the Ising model. 

At the special values $\phi = 0,\pi/2$ and $\pi$, corresponding to $b=\infty, 0$ and $-\infty$, respectively, the interfaces reduce to separate boundary conditions for the two Ising models given by

\begin{equation}
 (++)\oplus (--)\,,~~(ff)~~\text{and}~~(+-)\oplus(-+)\,,
\end{equation}
where $+,-,f$ dennote the three conformal boundary conditions of the Ising model, namely spin-up, spin-down and free \cite{cardy_boundary_1989}. 

For Neumann interfaces, or order-disorder interfaces, the relation between $\tilde\phi$ and $\tilde b$ is similar but with $\tilde\phi\in [0,\pi/2]$ and thus $\tilde b\ge0$ (see e.g. in \cite{bachas_fusion_2013}). Again, for the special value $\tilde \phi = \pi/4$ we have $\tilde b = 1$. This means that this Neumann interface is topological. On the other hand, at the values $\tilde \phi = 0,\pi/2$ the interfaces reduces to separate boundary conditions 

\begin{equation}
 (+f)\oplus(-f)~~\text{and}~~(f+)\oplus(f-)\,.
\end{equation}

Other two interesting quantities that characterise all conformal interfaces are the reflection coefficient $\mc{R}$ and the transmission coefficient $\mc{T}$ which are given by 2-point functions of the energy momentum tensor as follows \cite{quella_reflection_2007}

\begin{equation}
 \mc{R} \equiv \frac{\langle T_1 \bar{T}_1 +T_2\bar{T}_2\rangle}{\langle (T_1+\bar{T}_2)(\bar{T}_1+T_2) \rangle}\,,~~~~\mc{T} \equiv \frac{\langle T_1 \bar{T}_2 +T_2\bar{T}_1\rangle}{\langle (T_1+\bar{T}_2)(\bar{T}_1+T_2) \rangle} \label{eq:ReflTrans}
\end{equation}
where $T_1,\bar{T}_1$ are the components of the energy momentum tensor at the point $z$ and $T_2,\bar{T}_2$ are evaluated at the corresponding point reflected at the interface. For the interfaces we considered previously the reflection and transmission coefficients are given by

\begin{equation}
 \mc{R} = \left\{\begin{matrix}
                  \cos^2(2\phi) &\text{Dirichlet}\\
                  \cos^2(2\tilde\phi) &~\text{Neumann}
                 \end{matrix}\right.\,,~\text{and}~~~\mc{T} = \left\{\begin{matrix}
                  \sin^2(2\phi) &\text{Dirichlet}\\
                  \sin^2(2\tilde\phi) &~\text{Neumann}
                 \end{matrix}\right.\,.                 
\end{equation}

It is easy to see that $\mc{R}+\mc{T} = 1$. Note that for topological Dirichlet interfaces, where $\phi=\pi/4$ or $3\pi/4$, there is no reflection, namely $\mc{R} = 0$. On the other hand, for $\phi = n\pi/2$ the reflection coefficient is $\mc{R}=1$, and thus the interface reduces to a totally reflecting boundary conditions. For Neumann interfaces the statements are alike. 

\subsection{Free fermion description}
\label{sec:explicitInterfaces}

While the description in terms of the free boson provides an overview over the possible interfaces, to construct the explicit interface operator one needs to undo the folding. This is best done in the language of free fermions. Recall that the Ising model can be regarded as a system of a free real Majorana fermion, where  modular invariance is achieved by a projection on even fermion number (where the fermion number is the sum of left and right fermion number). In a free fermion theory one distinguishes between the NS-sector and the R-sector. In the NS sector, the fermions $\psi, \bar\psi$ (denoting left and rightmovers) are half integer moded and there is a non-degenerate ground state. In the R-sector the fermions are integer moded  and the ground state degenerates. The Ising model has three primary fields with respect to the Virasoro algebra, $1, \sigma, \epsilon$ of left-right conformal dimensions $(0,0), (1/16, 1/16), (1/2, 1/2)$. In terms of the free fermion $(0,0)$ is the NS-vacuum, $(1/2,1/2)$ the first excited state of the NS-sector and $(1/16, 1/16)$ a R-ground state (after the degeneracy of the ground states has been lifted by the GSO-projection).

Having an interface between two 2D free fermion conformal field theories as on the left of \fref{fig:foldingtrick}, its interface operator has the general form \cite{bachas_worldsheet_2012,bachas_fusion_2013}

\begin{equation}
 I_{1,2}(O) = \prod_{n>0} I_{1,2}^n(\mathcal{O}) I_{1,2}^0(O) \equiv I_{1,2}^>(\mathcal{O}) I_{1,2}^0(\mathcal{O})\,,
\end{equation}
where we have split the operator into two factors; $ I_{1,2}^0(\mathcal{O})$ is a map of the ground states of the free fermion theory, whereas $I_{1,2}^>$ contains the higher oscillator modes. The latter can be factorized further;  in $I_{1,2}^n$ only the $\pm n$th modes of the fermion field appear pairwise, such that all $I_{1,2}^n$ commute. It is given by 

\begin{equation}
 I^{n}_{1,2} = \text{exp} \left(-i\psi^1_{-n}\mathcal{O}_{11}\bar{\psi}^1_{-n}+\psi^1_{-n}\mathcal{O}_{12}\psi^2_n+\bar \psi^{1}_{-n}\mathcal{O}_{21}\bar \psi^{2}_{n}+i\psi^{2}_{n}\mathcal{O}_{22}\bar \psi^{2}_{n}\right), \label{eq:Iexp}
\end{equation}
where the $\psi^{1/2}_{\mp n}$ are the modes of CFT1/CFT2 which are acting from the left/right on $I^0$ -- the ground state operators. The matrix $\mc{O} \in O(2)$ specifies the interface and can be given in terms of a boost matrix $\Lambda\in O(1,1)$ which guarantees that the interface preserves conformal invariance, see \cite{bachas_worldsheet_2012,bachas_fusion_2013} for more details. Their exact relation is given by 

\begin{equation}
 \mathcal{O}(\Lambda) = \begin{pmatrix}
                         \Lambda_{12}\Lambda_{22}^{-1} & \Lambda_{11}- \Lambda_{12}\Lambda_{22}^{-1}\Lambda_{21} \\
                         \Lambda_{22}^{-1} & -\Lambda_{22}^{-1}\Lambda_{21}
                        \end{pmatrix}.
\end{equation}

The matrices with det$\Lambda = +1$ correspond to Dirichlet boundary conditions in the orbifold theory whereas det$\Lambda = -1$ corresponds to the Neumann boundary conditions. For $\det \Lambda = +1$ the  relation gives

\begin{equation}
 \Lambda = \begin{pmatrix}
            \cosh\gamma & \sinh\gamma\\
            \sinh\gamma & \cosh\gamma
           \end{pmatrix} ~ \leftrightarrow ~\mathcal{O} = \begin{pmatrix}
							 \cos(2\phi) & \sin(2\phi)\\
							 \sin(2\phi) & -\cos(2\phi)
						       \end{pmatrix},\label{eq:Ophi}
\end{equation}
and for $\det\Lambda=-1$

\begin{equation}
 \Lambda = \begin{pmatrix}
            \cosh\tilde\gamma & -\sinh\tilde\gamma\\
            \sinh\tilde\gamma & -\cosh\tilde\gamma
           \end{pmatrix} ~ \leftrightarrow~ \mathcal{O} = \begin{pmatrix}
							 \cos(2\tilde\phi) & \sin(2\tilde\phi)\\
							 -\sin(2\tilde\phi) & \cos(2\tilde\phi)
						       \end{pmatrix}\,.\label{eq:Ophi-}
\end{equation}
Indeed, $\phi$ and $\tilde \phi$ precisely correspond to the parameters describing the D0 and D1 brane moduli space.
From now on we omit the tildes. Then we can write $\cos(2\phi) = \tanh\gamma \Leftrightarrow e^\gamma = \cot\phi$ in both cases. To obtain the interface operators of the Ising model from those of the free fermion theory, one still has to GSO-project on total even fermion number. This requires taking linear combinations of the free fermion interfaces.  In the Ising model the type-0-GSO projection allows us to distinguish three cases: The interface operators for $\det\Lambda=1$ that carry either positive or negative RR charge and can be written as \footnote{ See \cite{bachas_fusion_2013} for more details on the construction}.

\begin{align}
 I^{\pm}(\Lambda) = \frac{1}{2} \left(I^{\text{NS}}(\Lambda) \pm I^\text{R}(\Lambda) \right) + (\Lambda \rightarrow -\Lambda)\,, \label{eq:chargedInterface}
\end{align}
and these for $\det\Lambda=-1$ which are the neutral operators 

\begin{equation}
 I^\text{n.}(\Lambda) = \frac{1}{\sqrt{2}}I^{\text{NS}}(\Lambda) + (\Lambda \rightarrow -\Lambda)\,. \label{eq:neutralInterface}
\end{equation}

The operators $I^{\text{NS}}$ and $I^\text{R}$ act on the Neveu-Schwarz and Ramond sector of the free fermion theory, respectively. They are given by 

\begin{align}
 I^{\text{NS}}(\Lambda) = \prod_{n\in\mathbb{N}-\frac{1}{2}} I^n(\Lambda) \,I^{0,\text{NS}}\,, ~~~\text{with}~~~ I^{0,\text{NS}} = \ket{0}_\text{NS}\,_\text{NS}\bra{0}\,,
\end{align}
and

\begin{equation}
\begin{split}
 I^{\text{R}}(\Lambda) &= \prod_{n\in\mathbb{N}} I^n(\Lambda) \,I^{0,\text{R}}\,,\\[1em] 
 ~~~\text{with}~~~~~ I^{0,\text{R}} &= \sqrt{\vert \sin(2\phi)\vert}\, \Big(\ket{+}_\text{R}\,_\text{R}\bra{+}+\ket{-}_\text{R}\,_\text{R}\bra{-}\Big) S(\Lambda)\,\\
 &= \sqrt2 \Big(\cos(\phi)\,\ket{+}_\text{R}\,_\text{R}\bra{+}+\sin(\phi)\,\ket{-}_\text{R}\,_\text{R}\bra{-}\Big) \,.
\end{split}
\end{equation}
Here, $\ket{\pm}_\text{R}$ denote R-ground states and $S$ is the spinor representation of $O(1,1)$.

\section{How to derive the entanglement entropy}
\label{sec:howtoEE}

In this section we briefly want to show the derivation of formulas we use in the following sections. It is an adaptation of the procedure used to derive the entanglement entropy through interfaces in the free boson theory. Thus, for a more detailed derivation we recommend \cite{sakai_entanglement_2008}, but also \cite{holzhey_geometric_1994}. 

Our goal is to derive the (ground state) entanglement entropy between two 2D GSO-projected free fermionic CFTs connected by the interfaces \eqref{eq:chargedInterface} or \eqref{eq:neutralInterface}. We formally define CFT1 to live on a half complex plane $\text{Re}\, w > 0$ and CFT2 on $\text{Re} \,w < 0$, respectively. The interface then lies on the imaginary axis $\text{Re}\, w = 0$. The EE is defined by the von Neumann entropy of the reduced density matrix for the ground state $\rho_1 = \Tr_2 \vert0\rangle\langle0\vert$ as (see e.g. \cite{calabrese_entanglement_2004,calabrese_entanglement_2009})

\begin{equation}
 S = -\text{Tr}_1\, \rho_1 \log \rho_1 \equiv -\partial_K \text{Tr}_1\,\rho_1^K \vert_{K\rightarrow1}\,.
\end{equation}

The trace of the $K$-th power of the reduced density matrix is also given by the partition function on a $K$-sheeted Riemann surface $\mathcal{R}_K$ with a branch cut along the positive real axis \cite{holzhey_geometric_1994,sakai_entanglement_2008}

\begin{equation}
 \text{Tr}_1 \rho_1^K = \frac{Z(K)}{Z(1)^K}\,. \label{eq:Z(K)/Z}
\end{equation}

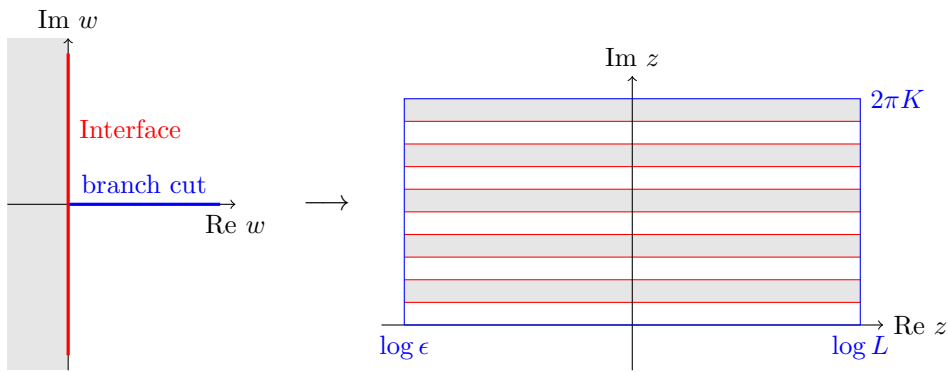
\begin{figure}[ht!]
 \begin{center}
  \begin{tikzpicture} [scale=2]
   \fill [gray!20] (-.4,1.1) rectangle (0,-1.1);
   \draw[->] (-.4,0) -- (1.1,0) node[below] {\small{Re $w$}};
   \draw[->] (0,-1.1) -- (0,1.1) node[above] {\small{Im $w$}};
   \draw[blue,very thick] (0,0) -- node[above] {\small{branch cut}} (1,0);
   \draw[very thick,red] (0,-1) -- (0,0);
   \draw[very thick,red] (0,0) -- node[right] {\small{Interface}} (0,1);
   \node at (1.7,0) {$\longrightarrow$};
  \end{tikzpicture}
  \begin{tikzpicture} [scale=3]
   \foreach \j in {1,...,5} \fill[gray!20] (-1,2*.\j) rectangle (1,2*.\j-.1);
   \foreach \i in {1,...,9} \draw[red] (-1,.\i)--(1,.\i);
   \draw[->] (-1.1,0) -- (1.1,0) node[right] {\small{Re $z$}};
   \draw[->] (0,-.2) -- (0,1.1) node[above] {\small{Im $z$}};
   \draw[blue] (-1,0) node[below] {\small{$\log \epsilon$}} --  (1,0) node[below] {\small{$\log L$}} -- (1,1) node[right] {\small{$2\pi K$}}-- (-1,1) -- (-1,0);
  \end{tikzpicture}
 \end{center}
 \caption{Sketch of the $K$-sheet Riemann surface $\mc{R}_K$ that one needs for the replica trick. After imposing a IR citoff $\epsilon$, an UV cutoff $L$, and suitable variable transformation, $\mc{R}_K$ looks like on the right. }
 \label{fig:InterfaceAndSheet}
\end{figure}

The $K$-sheet construction is illustrated on the left of \fref{fig:InterfaceAndSheet}. This procedure is a version of the so-called \textit{replica trick}. By the use of \eqref{eq:Z(K)/Z} the entanglement entropy can be written as

\begin{equation}
 S = (1-\partial_K)\log Z(K) \vert_{K\rightarrow1}\,. \label{eq:masterformula1}
\end{equation}

To evaluate the partition function $Z(K)$ we change coordinates to $z=\log w$ and introduce the cutoffs $\vert w \vert =\epsilon$ and $\vert w \vert =L$. The $K$-sheet $\mathcal{R}_K$ then looks as on the left part in \fref{fig:InterfaceAndSheet}. As in \cite{sakai_entanglement_2008} we impose periodic boundary conditions in Re $z$ and choose $\epsilon = \frac{1}{L}$ for simplicity. We then end up with the torus partition function with $2K$ interfaces inserted which is given by

\begin{align}
 Z(K) &= \text{Tr}_1 \left(I_{1,2} \,e^{-\delta H_2} \,I_{2,1} \,e^{-\delta H_1} \cdots I_{2,1} \,e^{-\delta H_1} \right) \nonumber \\
  &= \text{Tr}_1 \Big(I_{1,2} \,e^{-\delta H_2}\, I_{1,2}^\dagger \,e^{-\delta H_1} \Big)^K \,,\label{eq:partFct1}
\end{align}
with $\delta = \pi^2/(\log L)$ and $I_{2,1}=I_{1,2}^\dagger$. 

\section{Derivation of the partition function}
\label{sec:derPartition}

In the following we explicitly derive the partition function \eqref{eq:partFct1} for the interface operators introduced in section \ref{sec:explicitInterfaces}. We start with a single NS operator which is the simplest case. Step by step we show how to derive $Z(K)$ for the more complicated R, neutral, and charged operator. 

\begin{boldmath}
\subsection{The partition function for a single NS operator \texorpdfstring{$I = I^\text{NS}(\Lambda)$}{I}}
\end{boldmath}\label{sec:singleNS}
\noindent
In the NS sector of the free fermion theory we can formally write the Hilbert space $\mc{H}$ of the theory as a tensor product $\otimes_n \mc{H}_n$ where $\mc{H}_n = \text{span}\left\{\ket{0},\psi_{-n}\bar\psi_{-n}\ket{0},\psi_{-n}\ket{0},\bar\psi_{-n}\ket{0}\right\}$. Then each $I^n(\mc{O})$ as in \eqref{eq:Iexp} has a matrix representation in $\mc{H}_n$ given by

\begin{equation}
 I^n = \begin{pmatrix}
        1& - i\, \mc{O}_{22}&0&0\\
        -i \,\mc{O}_{11}&-\det\mc{O}&0&0\\
        0&0&\mc{O}_{12}&0\\
        0&0&0&\mc{O}_{21}\\
       \end{pmatrix}\,,
\end{equation}
and we can write $I^\text{NS} = \otimes_n I^n$. In this notation the propagator is given by $e^{-\delta H}\equiv P = \otimes_n P^n$ with

\begin{equation}
 P^n = \text{diag}(1,e^{-2\delta n},e^{-\delta n},e^{-\delta n})\,.
\end{equation}

Using the above notation with CFT1 = CFT2 the partition function \eqref{eq:partFct1} of the $K$-sheet can be written as

\begin{align}
 Z(K) &= \prod_{n\in \mathbb{N}-\frac{1}{2}} \Big(I^n\,P^n\,(I^n)^\dagger P^n \Big)^K \\
      &= \prod_{n\in \mathbb{N}-\frac{1}{2}} \left(\lambda_{n,1}^K +\lambda_{n,2}^K +\lambda_{n,3}^K +\lambda_{n,4}^K\right)\,,
\end{align}
where $\lambda_{n,i}$, $i=1,..,4$, are the eigenvalues of $D^n = \left[I^n\,P^n\,(I^n)^\dagger P^n\right]$.

\subsubsection{Explicit calculation}

\label{sec:explNS}

We have two distinguishable interfaces: the Dirichlet-interface with $\det\Lambda = 1$ and the Neumann-interface with $\det\Lambda=-1$. However, in both cases the matrices $D^n$ are similar and their eigenvalues are given by

\begin{align*}
 \lambda_{n,1} &\equiv e^{-2 n \delta}p^+_n = e^{-2 n \delta}\left(\cosh(2 n \delta) + \cos^2(2\phi) + \cosh(n\delta) \sqrt{2 \cosh(2n\delta)+2 \cos(4\phi)}\right)\,,\\[1em]
 \lambda_{n,2} &\equiv e^{-2 n \delta} p^-_n = e^{-2 n \delta}\left(\cosh(2 n \delta) + \cos^2(2\phi) - \cosh(n\delta) \sqrt{2 \cosh(2n\delta)+2 \cos(4\phi)}\right)\,,\\[1em]
 \lambda_{n,3} &= e^{-2n\delta} \sin^2(2\phi) =\lambda_{n,4}\,,
\end{align*}
so that the partition function of the $K$-sheet is given by 

\begin{equation}
  Z(K) = \prod_{n\in \mathbb{N}-\frac{1}{2}}e^{-2 K n \delta} \left(2 \sin^{2K}(2 \phi) +(p^+_n)^K+(p^-_n)^K\right)\,. \label{eq:R3+}
\end{equation}

At this stage one could proceed further by directly using formula \eqref{eq:masterformula1} on the latter result for $Z(K)$. Through the logarithm the infinite product simplifies to a sum. Taking the derivative w.r.t. $K$ in every summand it is then easy to write down a result for the entanglement entropy by means of an infinite sum. One could then evaluate the sum -- and thus the entanglement entropy -- numerically for every $\delta>0$ and $\phi$ up to arbitrary accuracy. However, we are mainly interested in  small $\delta$ -- which means large $L$ -- behaviour of the entanglement entropy, since $L$ is introduced as a UV cutoff. In this limit we can derive the EE analytically by proceeding as in the following.  

For odd $K$ the partition function \eqref{eq:R3+} can be written as

\begin{equation}
 Z(K) = \prod_{n\in \mathbb{N}-\frac{1}{2}}\left(\prod_{k=1}^K  2 e^{-2 n \delta } \left(2 \cos^2(\nu_k) -1 + \cosh(2 n \delta ) \right)\right)\,, \label{eq:oddZ}
\end{equation}
with $\nu_k = \arcsin\left(\sin(2\phi) \vert\sin(\frac{k \pi}{K})\vert\right)$. For even $K$ we have to add ~$4 e^{-2 K n \delta} \sin^{2K}(2\phi)$ ~to every factor in \eqref{eq:oddZ}.  Additionally, we state that the fraction $\theta[0,0](\tau,z)/\eta(\tau)$ of the well known $\theta$-function and $\eta$-functions as defined in \eqref{eq:eta} and \eqref{eq:genTheta} can be written as

\begin{equation}
 \frac{\theta[0,0]\hspace{-1pt}(\tau,z)}{\eta(\tau)} = e^{\frac{\pi i \tau}{12}}\prod_{n\in \mathbb{N}-\frac{1}{2}}2 e^{-2 n \pi i \tau} \left(2 \cos^2(\pi z) -1 + \cos\left(2 n \pi \tau\right) \right)\,. \label{eq:nthFactor}
\end{equation}

Thus we can conclude that the $K$-sheet partition function for odd $K$ can be expressed as

\begin{equation}
 Z(K) = e^{-\frac{K \delta}{12}}\,\prod_{k=1}^K \frac{\theta[0,0]\hspace{-2pt}\left(\frac{i \delta}{\pi} ,\frac{\nu_k}{\pi}\right)}{\eta\hspace{-1pt}\left(\frac{ i \delta}{\pi} \right)} \,. 
\end{equation}

Using the behaviour of $\eta$ and $\theta$ under $S$-transformations we can write

\begin{equation}
\begin{split}
 &\,Z(K) = e^{-\frac{K \delta}{12}}  \prod_{k=1}^K e^{-\frac{\nu_k^2}{\delta}}\,\frac{\theta[0,0]\hspace{-2pt}\left(i\frac{\pi}{\delta},-i\frac{\nu_k}{\delta}\right)}{\eta\hspace{-2pt}\left(i \frac{\pi}{\delta}\right)} \\[1em]
\xrightarrow{\delta \ll 1} ~&\boxed{Z(K) = \,e^{\frac{\pi^2K}{12 \,\delta}} \,e^{-\frac{\varphi(K)}{\delta}} \left(1+e^{-\frac{\mu}{\delta}}\right)}\,,
\end{split}\label{eq:ZKNS}
\end{equation}
where $\varphi(K) = \sum_{k=1}^K\nu_k^2$, and $\mu$ is constant in $\delta$.

For even $K$, $Z(K)$ can not be given in terms of $\theta$ and $\eta$ as above. One might wonder if we really can use \eqref{eq:ZKNS} to calculate the entanglement entropy although it is just valid for odd $K$. In Appendix \ref{sec:ArgJustOdd} we actually show why it really suffices to consider \eqref{eq:ZKNS}. 

\begin{boldmath}
\subsection{The partition function for a single R-operator \texorpdfstring{$I = I^\text{R}(\Lambda)$}{I}}
\end{boldmath}
\noindent
In the Ramond sector the modes are integer and the zero-mode map is slightly more difficult, $I^{0,\text{R}}  = \sqrt2 \Big(\cos(\phi)\,\ket{+}_\text{R}\,_\text{R}\bra{+}+\sin(\phi)\,\ket{-}_\text{R}\,_\text{R}\bra{-}\Big)$. The latter allows us to write 

\begin{equation}
 I^\text{R} = \cos(\phi)\,I^\text{R}_+ +\sin(\phi)\, I^\text{R}_-\,, ~~~ \text{with}~~~ I^\text{R}_\pm = \sqrt2 \, \ket{\pm}\bra{\pm} \prod_{n\in \mathbb{N}} I^n\,,
\end{equation}
where $I^\text{R}_-(\Lambda)\cdot I^\text{R}_+(\Lambda')$ vanishes. Proceeding similar to the case of NS-operators one gets

\begin{align}
 Z(K) 
 &= 2^{K} \,(\cos(\phi)^{2K}+\sin(\phi)^{2K})\prod_{n\in \mathbb{N}} \left(\lambda_{n,1}^K +\lambda_{n,2}^K +\lambda_{n,3}^K +\lambda_{n,4}^K\right)\,,
\end{align}
where again $\lambda_{n,i}$, $i=1,..,4$, are the eigenvalues of $\left[I^n\,P^n\,(I^n)^\dagger P^n\right]$.
\medskip

\subsubsection{Explicit calculation}

The eigenvalues for the R-interface are similar to the eigenvalues for the NS interface but with $n \in \mathbb{N}$. Thus, for odd $K$ we can write 

\begin{equation}
 Z(K) = 2^{K} \,\left(\cos(\phi)^{2K}+\sin(\phi)^{2K}\right) \prod_{k=1}^K\left(\prod_{n\in\mathbb{N}}2 e^{-2 \delta n} \left(2 \cos^2(\nu_k) -1 + \cosh(2 \delta n) \right)\right)\,,
\end{equation}
where again $\nu_k = \arcsin\left(\sin(2\phi) \vert\sin(\frac{k \pi}{K})\vert\right)$. This time the latter is given in terms of $\theta[\frac{1}{2},0](\tau,z)/\eta(\tau)$ because of $n$ being integer. One important difference between the $\theta$-function we use here and the $\theta$-function used in the case of the NS-interface is that there appears an additional factor of $2 \cos(\pi z)$. One can show that $$\prod_{k=1}^K \cos(\nu_k) = \cos(\phi)^{2K}+\sin(\phi)^{2K}$$ for odd $K$ so that in this case the partition function reduces to

\begin{equation}
\begin{split}
 Z(K) &=  e^{-\frac{K\delta}{6}} \,(\cos(\phi)^{2K}+\sin(\phi)^{2K})\, \prod_{k=1}^K \frac{1}{\cos(\nu_k)} \,\prod_{k=1}^K \frac{\theta[\frac{1}{2},0]\left(\frac{i \delta}{\pi} ,\frac{\nu_k}{\pi}\right)}{\eta\left(\frac{ i \delta}{\pi} \right)}\\
 &= e^{-\frac{K\delta}{6}} \,\prod_{k=1}^K \frac{\theta[\frac{1}{2},0]\left(\frac{i \delta}{\pi} ,\frac{\nu_k}{\pi}\right)}{\eta\left(\frac{ i \delta}{\pi} \right)}\,.
\end{split}
\end{equation}

The same steps as for the NS interface now lead us to

\begin{equation}
 \boxed{Z(K)  = e^{\frac{\pi^2}{12 \,\delta}K}\,e^{-\frac{\varphi(K)}{\delta}} \left(1+e^{-\frac{\mu}{\delta}}\right)}\,, \label{eq:ZKR}
\end{equation}
in the limit $\delta\ll 1$.

In the Ramond sector only the Dirichlet interfaces have non-trivial components, so we do not consider Neumann boundary conditions, although they would not make any difference for $Z(K)$. 

\subsection{The partition function for the neutral interface operator}

The neutral interface operator is given by $I^\text{n.}(\Lambda) = \frac{1}{\sqrt2}\left(I^\text{NS}(\Lambda) + I^\text{NS}(-\Lambda)\right)$ with Neumann boundary conditions, i.e. $\det \Lambda = -1$. Some simple algebra leads to

\begin{align}
 2 \,I^\text{n.}(\Lambda)\, e^{-\delta H} \,I^{\text{n.}}(\Lambda)^\dagger \,e^{-\delta H} =&~~~~ I^\text{NS}(\Lambda) \,e^{-\delta H}\, I^\text{NS}(\Lambda) \,e^{-\delta H} + I^\text{NS}(-\Lambda) \,e^{-\delta H}\, I^\text{NS}(-\Lambda) \,e^{-\delta H} \nonumber\\
&+ I^\text{NS}(-\Lambda) \,e^{-\delta H}\, I^\text{NS}(\Lambda) \,e^{-\delta H} + I^\text{NS}(\Lambda) \,e^{-\delta H}\, I^\text{NS}(-\Lambda) \,e^{-\delta H}\nonumber\\[.5em]
= &~~~\, 2 \left( I^\text{NS}(\Lambda) \,e^{-\delta H}\, I^\text{NS}(\Lambda) \,e^{-\delta H} \right)\\
&+ 2 \left( I^\text{NS}(\Lambda) \,e^{-\delta H}\, I^\text{NS}(-\Lambda) \,e^{-\delta H} \right) \nonumber\\[.5em]
\equiv &~ 2 \left(D_+ + D_- \right)\,,\nonumber
\end{align}
with $D_\pm$ given by the tensor product of the matrices

\begin{equation}
D_\pm^n =\begin{pmatrix}
                     1+ e^{-2\delta n} \,\mc{R} & i(e^{-4\delta n}+e^{-6\delta n})\sqrt{\mc{R}}  &0&0\\
                     -i(e^{-2\delta n}+e^{-4\delta n})\sqrt{\mc{R}}&e^{-4\delta n}+ e^{-2\delta n} \,\mc{R}&0&0\\
                     0&0&\pm e^{-2\delta n} \,\mc{T}&0\\
                     0&0&0&\pm e^{-2\delta n} \,\mc{T}
\end{pmatrix}\,,\label{eq:matrixD}
\end{equation}
where we here used the reflection coefficient $\mc{R} = \cos(2\phi)^2$ and the transmission coefficient $\mc{T} = \sin(2\phi)^2$ as introduced in \eqref{eq:ReflTrans}. We can see that both $D_+$ and $D_-$ can be diagonalized simultaneously. With straight forward linear algebra we can now calculate the partition function for the $K$-sheet with a neutral interface insertion. It is given by

\begin{equation}
\begin{split}
  Z(K) &= \Tr\left(D_+ + D_- \right)^K\\[.5em]
    &=~~2^{K-1}\prod_{n\in \mathbb{N}-\frac{1}{2}}e^{-2 K n \delta} \left(+2 \sin^{2K}(2 \phi) +(p^+_n)^K+(p^-_n)^K\right) +\\
   &~~~+ 2^{K-1}\prod_{n\in \mathbb{N}-\frac{1}{2}}e^{-2 K n \delta} \left(-2 \sin^{2K}(2 \phi) +(p^+_n)^K+(p^-_n)^K\right)\,.
\end{split} \label{eq:Zneutral}
\end{equation}

The important factor of $2^{K-1}$ can be understood with the following simpler example: Consider the matrices $M_\pm(x) = \otimes_i\text{diag}(x,\pm x)$. It is now easy to convince oneself that $$[M_+(x) +M_-(x)]\cdot[M_+(y) +M_-(y)] = 2 [M_+(xy) +M_-(xy)],$$ which allows us to directly conclude that $[M_+(x) +M_-(x)]^K = 2^{K-1} [M_+(x^K) +M_-(x^K)]$. The generalization to $D_+$ and $D_-$ is straight forward. 

The first summand in \eqref{eq:Zneutral} is the same as for the single NS-interface. In a very similar way as in section \ref{sec:singleNS}, the second summand can also be written in terms of $\theta[0,0]/\eta$. We here only state the result in the limit $\delta\ll1$:

\begin{equation}
 \boxed{Z(K) = 2^{K-1}e^{\frac{\pi^2\,K}{12 \,\delta}} \,\left( e^{-\frac{\varphi(K)}{\delta}} + e^{-\frac{\chi(K)}{\delta}}\right)}\,, \label{eq:ZKn}
\end{equation}
where $~\chi(K) = \sum_{k=1}^K \mu_k^2~$ with $~\mu_k = \arcsin(\sin(2\phi)\vert\cos(\frac{k\pi}{K})\vert)$\,.

\subsection{The partition function for the charged interface operator}

The charged interface operator is given by $I^{\pm}(\Lambda) = \frac{1}{2} \left(I^{\text{NS}}(\Lambda) \pm I^\text{R}(\Lambda) \right) + (\Lambda \rightarrow -\Lambda)$ with Dirichlet boundary conditions, i.e. $\det\Lambda= 1$. With similar considerations as for the neutral interface we can write 

\begin{equation}
\begin{split}
 2I^\pm(\Lambda)\, e^{-\delta H} \,I^\pm(\Lambda)^\dagger \,e^{-\delta H} &= \left(D_+ + D_-\right) \,\oplus 2\,\cos^2(\phi) \,\left(D_+^\text{R+}+D_-^\text{R+}\right) \,\oplus\\&~~~\oplus 2\,\sin^2(\phi) \,\left(D_+^{\text{R}-}+D_-^{\text{R}-}\right)\,, 
\end{split}
\end{equation}
where $D_\pm^\text{R+}$ corresponds to the vacuum $\ket{+}$ and $D_\pm^{\text{R}-}$ corresponds to $\ket{-}$ in a similar way to the single Ramond interface. There is no difference between the positively and the negatively charged interface. In matrix representation all the $D$'s are tensor products of matrices similar to \eqref{eq:matrixD} but with integers for the Ramond operators. Thus the partition function of the $K$-sheet with a charged interface can be written as

\begin{align}
  Z(K) &= \frac{1}{2^K}\Tr\Big[\left(D_+ + D_- \right)^K  \oplus 2^{K} \cos(\phi)^{2K} \left(D^\text{R+}_+ + D^\text{R+}_- \right)^K\oplus\nonumber\\
  &~~~~~~~~\oplus 2^{K} \sin(\phi)^{2K} \left(D^{\text{R}-}_+ + D^{\text{R}-}_- \right)^K\Big] \nonumber\\[.5em]
    &=\frac{1}{2} \bigg(\prod_{n\in \mathbb{N}-\frac{1}{2}}e^{-2 K n \delta} \left(+2 \sin^{2K}(2 \phi) +(p^+)^K+(p^-)^K\right) +\\
   &~~~+ \prod_{n\in \mathbb{N}-\frac{1}{2}}e^{-2 K n \delta} \left(-2 \sin^{2K}(2 \phi) +(p^+)^K+(p^-)^K\right)+\nonumber\\
   &~~~+ 2^{K} \left(\cos(\phi)^{2K} +  \sin(\phi)^{2K}\right)  \prod_{n\in \mathbb{N}} e^{-2 K n \delta} \left(+2 \sin^{2K}(2 \phi) +(p^+)^K+(p^-)^K\right)+\nonumber\\
   &~~~+ 2^{K} \left(\cos(\phi)^{2K} +  \sin(\phi)^{2K}\right)  \prod_{n\in \mathbb{N}} e^{-2 K n \delta} \left(-2 \sin^{2K}(2 \phi) +(p^+)^K+(p^-)^K\right)\bigg).\nonumber
\end{align}

Using the same logic as in the previous sections, the partition function for odd $K$ reduces to

\begin{equation}
 \boxed{Z(K) = e^{\frac{\pi^2\,K}{12 \,\delta}} \left( e^{-\frac{\varphi(K)}{\delta}} + f(K)\,e^{-\frac{\chi(K)}{\delta}}\right)}\,, \label{eq:ZKc}
\end{equation}
in the limit $\delta\ll 1$. The functions $\phi$ and $\chi$ are given as before and $f(K)$ is given by

\begin{equation}
 f(K) = \frac{1}{2}\left(1+ \left(\cos(\phi)^{2K} +  \sin(\phi)^{2K}\right) \prod_{k=1}^K \frac{1}{\cos(\mu_k)} \right) \neq 1\,.
\end{equation}

\section{Derivation of the entanglement entropy}
\label{sec:derEE}
Before we explicitly derive the entanglement entropy we want to show that -- for $\delta\ll 1$ -- it is the same in all previous cases up to an additional $\log2$ term for the neutral interface. Our formula of choice \eqref{eq:masterformula1} is 

\begin{equation}
 S = (1-\partial_K) \log Z(K) \vert_{K\rightarrow1}\,, \label{eq:masterformula}
\end{equation}
for which it is easy to check that any overall factor $C^K$ in $Z(K)$ with $C$ constant in $K$ does not contribute to the entanglement entropy. 

At first we want to write down the preliminary result for the single NS-interface where $Z(K)$ is given by \eqref{eq:ZKNS} so that the entanglement entropy can be written as

\begin{equation}
 S^\text{NS} = \left(- \varphi(1) + \partial_K \varphi (1)\right) \frac{1}{\delta}\,.
\end{equation}

Next we consider the single R-interface where the partition function is given by \eqref{eq:ZKR}. Its EE is simply the same as for the NS-interface

\begin{equation}
 \boxed{S^\text{R} = \left(- \varphi(1) + \partial_K \varphi (1)\right) \frac{1}{\delta} = S^\text{NS}}\,.
\end{equation}


Now we want to derive the EE for the neutral interface. Inserting its partition function \eqref{eq:ZKn} in \eqref{eq:masterformula} gives 

\begin{equation}
\begin{split}
 &\,S^\text{n.} = \log\left(e^{-\varphi(1)/\delta}+e^{-\chi(1)/\delta}\right) + \frac{1}{\delta}\,\frac{\partial_K \varphi(1)\,e^{-\varphi(1)/\delta}+\partial_K\chi(1)\,e^{-\chi(1)/\delta}}{e^{-\varphi(1)/\delta}+e^{-\chi(1)/\delta}}-\log2\\
&~~~~\,= \left(-\varphi(1) + \partial_K \varphi(1)\right)\frac{1}{\delta} -\log2\\ 
 &\boxed{S^\text{n.}  = S^\text{NS}-\log2}\,,
\end{split}
\end{equation}
where we can simplify to the second line because we are in the limit $\delta \ll 1$ and because $\varphi(1) = 0 > \chi(1) = -4\phi^2$. 

At last we want to derive the EE for the charged interface operator where $Z(K)$ is given by \eqref{eq:ZKc}. As in the case of the neutral interface operator every term with a factor $e^{-\chi(1)/\delta}$ can be neglected. Consequently also $f(K)$ in \eqref{eq:ZKc} has no contribution to the entanglement entropy when we are in the limit $\delta\ll1$. It again simply reduces to the EE for the single NS-interface: 

\begin{equation}
 \boxed{S^\pm = S^\text{NS}}\,.
\end{equation}

\begin{boldmath}
\subsection{Explicit derivation of \texorpdfstring{$S^\text{NS}$}{S}}
\end{boldmath}
\noindent
To derive the entanglement entropy explicitly we have to calculate $\partial_K\varphi(1)$. Therefore we proceed similar as in \cite{sakai_entanglement_2008} and write $\varphi(K) = \sum_{k=1}^K\nu_k^2 \equiv \sum_{k=1}^K f\left(\frac{k}{K}\right)$, which can be written as a Taylor series around $k/K = 0$ and further massaged as 

\begin{equation}
\begin{split}
 \varphi(K) &= \sum_{k=1}^K \sum_{m=0}^\infty f_m \left(\frac{k}{K}\right)^m= \sum_{m=0}^\infty \frac{f_m}{K^m} \sum_{k=1}^K k^m\\
  &= \sum_{m=0}^\infty \frac{f_m}{K^m}\, \frac{B_{m+1}(K+1)-B_{m+1}}{m+1} \,,
\end{split}
\end{equation}
where $B_{n}(x),B_{n}$ are the Bernoulli polynomials and Bernoulli numbers, respectively, as given in Appendix \ref{sec:Bernoulli}. Its derivative in the limit $K\rightarrow 1$ is then given by

\begin{equation}
\begin{split}
 \partial_K \varphi(K)\vert_{K\rightarrow 1} &= \sum_m \frac{f_m}{m+1} \, \underbrace{\partial_KB_{m+1}(K+1)}_{=\partial_K ( B_{m+1}(K) + (m+1)K^m)}\vert_{K\rightarrow 1} - \frac{f_m\,m}{m+1} \underbrace{(B_{m+1}(2) - B_{m+1})}_{=(m+1)} \\
 &= \sum_m \frac{f_m}{m+1} \,\partial_K B_{m+1}(K)\vert_{K\rightarrow 1} +f_m \,m -f_m \,m \\
 &= \sum_m \frac{f_m}{m+1} \,\partial_K B_{m+1}(K)\vert_{K\rightarrow 1}\,.
\end{split}\label{eq:dphi}
\end{equation}

At this stage we use the formula \eqref{eq:BernoulliDerivative} to obtain

\begin{equation}
 \partial_K \varphi(K)\vert_{K\rightarrow1} = f(0) +\frac{1}{2} f'(0) + \int_0^\infty \frac{i f'(it) - i f'(-it)}{1-e^{2\pi t}} dt\,.
\end{equation}
Both $f(0)$ and $f'(0)$ vanish, so that the entanglement entropy is given by

\IBox{\begin{equation}
 \begin{split}
  S^\text{NS} &= \frac{1}{\delta} \,\int_0^\infty \frac{i f'(it) - i f'(-it)}{1-e^{2\pi t}} dt \\
  &\equiv \frac{\pi^2 \sigma(s)}{\delta} = \sigma(s) \log(L)\,, 
 \end{split} \label{eq:SNS}
\end{equation}}
where we defined $s=|\sin(2\phi)| = \sqrt{\mc{T}}$ and 

\begin{equation}
\begin{split}
 \sigma(s) &= \frac{2}{\pi} \int_0^\infty \frac{\text{arcsinh}\left(s\,\sinh(\pi t)\right)\, \Big(\text{coth}(\pi t)-1\Big)\, s\,\cosh(\pi t)}{\sqrt{1+s^2\,\sinh^2(\pi t)}} dt \,\\
 & = \frac{2}{\pi^2} \int_0^\infty u \left(\sqrt{1+\frac{s^2}{\sinh^2(u)}}-1\right) du\,.
 \end{split} \label{eq:factor}
\end{equation}
For the last step we substituted $u = \text{arcsinh}\left(s\,\sinh(\pi t)\right)$.

\subsection{Comments on the result and special cases}

\begin{figure}[ht!]
\centering
 \includegraphics[width=.6\textwidth]{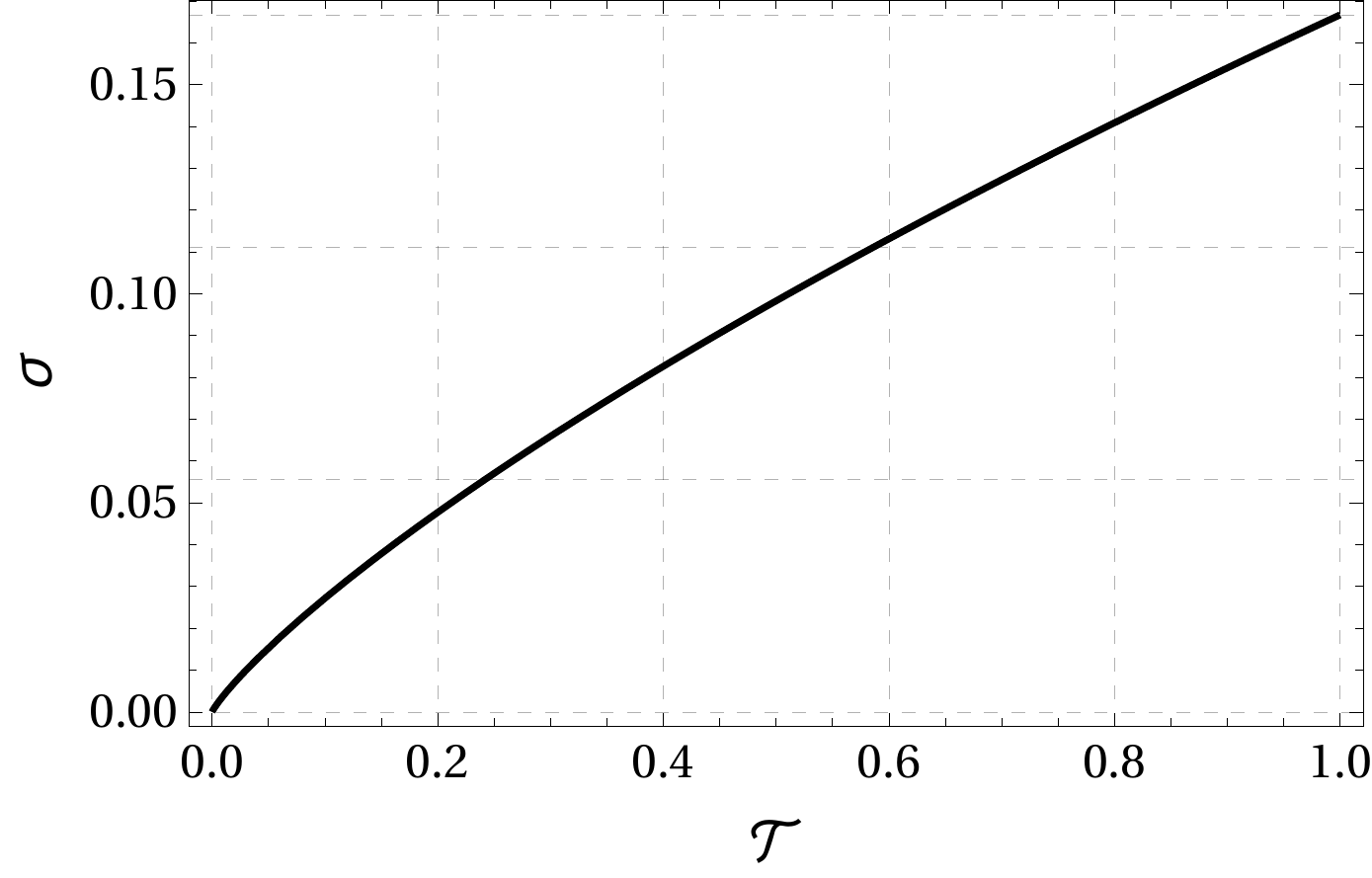}
 \caption{The prefactor $\sigma$ as a function of the transmission coefficient $\mc{T}$. As expected, it vanishes for totally reflecting interfaces, where $\mc{T} = 0$, and reproduces the familiar result $\sigma(\mc{T}=1) = 1/6 = c/3$ for topological interfaces in the Ising model.}
 \label{fig:sigma}
\end{figure}

The result \eqref{eq:SNS} shows that the EE through a single NS interface has a logarithmic scaling with respect to the (large) size $L$ of the system. Up to an additional term ``$\log2$'' for the neutral interface operator this is exactly the behaviour of the EE through a conformal interface in the 2D Ising model, too. The interface itself affects the EE mainly through the factor $\sigma(s)$ which is given in integral form in \eqref{eq:factor}. The square of the ``variable'' $s=|\sin(2\phi)|$ just is the transmission coefficient of the interface $\mc{T}$, but can be given in terms of the scaling factor $b$ and the coupling constant $K_1$ of the Ising model as in \eqref{eq:phi-b}, too. In \fref{fig:sigma}, we show the explicit dependence of the factor $\sigma$ on the transmission coefficient $\mc{T}$. 

There are two special cases to mention. First, we want to consider topological interfaces where the transmission coefficient is maximal,  $\mc{T} =1$. This case includes the identity in the free fermionic CFT and in the Ising model which is why it should reproduce the universal scaling of the entanglement entropy without interfaces. In fact, with $\sigma(1) = 1/6 = c/3$ we obtain the right result from \cite{calabrese_entanglement_2009}. However, not just the universal scaling is correct, but also the sub-leading term, i.e. contributions constant in $L$, are right. This can be checked by directly evaluating the torus partition function in the theory without interface, which is e.g. for the Ising model given by

\begin{equation}
 Z^\text{Ising}(K) = \frac{1}{2}\left(\frac{\theta[0,\frac{1}{2}]\left(\tau,0\right)}{\eta\left(\tau\right)}+\frac{\theta[\frac{1}{2},\frac{1}{2}]\left(\tau,0\right)}{\eta\left(\tau\right)}+\frac{\theta[\frac{1}{2},0]\left(\tau,0\right)}{\eta\left(\tau\right)}\right)\,,
\end{equation}
with $\tau = i\frac{\delta K}{\pi}$.

Secondly, in the limit of totally reflecting interface, when $\mc{T} \rightarrow 0$, one can show that $\sigma(\mc{T}) = \mc{O}(\mc{T}\log\sqrt{T})$, so that for vanishing transmittance  also the entanglement entropy vanishes. This fits the fact that all the oscillator parts of the two CFTs are decoupling for an interface with $\mc{T}\rightarrow0$. As shown in \fref{fig:sigma}, $\sigma(\mc{T})$ increases monotonically between these two extremal cases. The latter observation supports the intuitive guess that entanglement changes according to the transmittance of the interface. The lower the transmittance the lower the strength of interaction between the two CFTs connected by the interface. 

We also want to discuss the sub-leading term although it vanishes but for the neutral interface operator. An important contribution to that term for the Ramond and the charged interface comes from the ground state map.\footnote{Especially for bosonic interfaces this is often called the \textit{lattice part} of the interface operator. This name comes from the change of the lattice structure of the torus by the ground state mapping.} In our cases the ground state map is rather easy, it just separates the Hilbert space of the CFT in direct sums. Let us consider a Hilbert space $\mc{H} = \mc{H}_0 \oplus \mc{H}_1$ and two density operators $\rho_0$ and  $\rho_1$ acting on the respective Hilbert space, such that 

\begin{equation}
 \rho = \alpha \,\rho_0 \oplus \beta\, \rho_1\,, 
\end{equation}
where $\alpha$ and $\beta$ are real numbers. It is then easy to see that the von Neumann entropy on the full Hilbert space $S_\rho$ can be given in terms of the entropies on $\mc{H}_0$ and $\mc{H}_1$:

\begin{equation}
 S_\rho = \alpha \,S_{\rho_0} + \beta\, S_{\rho_1} - (\alpha \log\alpha + \beta \log\beta)\,.
\end{equation}

As an example let us consider the single Ramond interface. Because of the ground state map the (reduced) density matrix can be written as $\rho = \cos^{2}\hspace{-3pt}\phi \,\rho_+ + \sin^{2}\hspace{-3pt}\phi\, \rho_-$ and since $S_{\rho_+} = S_{\rho_-}$ the von Neumann entropy reads 

\begin{equation}
 S^\text{R} = S_\rho =  S_{\rho_+} - (\cos^2\hspace{-3pt}\phi\,\log\cos^2\hspace{-3pt}\phi + \sin^2\hspace{-3pt}\phi\,\log\sin^2\hspace{-3pt}\phi)\,.
\end{equation}

However, $S_{\rho_+}$ itself has an additional sub-leading term that exactly cancels the contribution of the ground state map. 

Another contribution to the sub-leading term comes from the GSO-projection which separates the Hilbert space of the free fermion in a direct sum $\mc{H} = \mc{H}_0\oplus\mc{H}_1$, graded by the fermion number. As an example consider the interface operator $I^\text{NS}(\Lambda)$ on the full NS Hilbert space. Then the projected interfaces on $\mc{H}_0$ and $\mc{H}_1$ are

\begin{equation}
 I_\pm = \frac{1}{\sqrt 2} (I^\text{NS}(\Lambda) \pm I^\text{NS}(-\Lambda))\,,
\end{equation}
where $I_+$ is the previously called neutral interface operator. A density operator is given by $I^\dagger I$, such that $\rho = \frac{1}{2} (\rho_+ + \rho_-)$ and thus

\begin{equation}
 S^\text{NS} = S_\rho = \frac{1}{2} (S_+ + S_-) + \log2\,,
\end{equation}
where in the limit of large $L$ it can be shown that $S_+ = S_-$, so that we get $S^\text{NS} = S^\text{n.} + \log2$ or 

\begin{equation}
 S^\text{n.} = S^\text{NS} - \log2\,.
\end{equation}

It is noteworthy that the sub-leading term does not depend on the most significant property of the interfaces, namely the transmittance $\mc{T}$, in all our cases. A similar result is known for entanglement entropy through interfaces in the free boson theory \cite{sakai_entanglement_2008}. There the sub-leading term only depends on the winding numbers $k_1,k_2$ on both sides of the interface and is simply given by ``$-\log |k_1k_2|$''. 

As a final comment in this section we want to state that one can also express the pre-factor $\sigma$ in terms of the Logarithm and Dilogarithm $\text{Li}_2$ in a similar way as in \cite{sakai_entanglement_2008}. It then looks like
\begin{equation}
 \sigma(s) = \frac{s}{6} - \frac{1}{6} -\frac{1}{\pi^2}\big((s+1)\log(s+1)\log s+(s-1)\text{Li}_2(1-s) + (s+1)\text{Li}_2(-s)\big)\,.
\end{equation}
This indeed agrees with formulas (22-26) of \cite{2010arXiv1005.2144E}.\footnote{In our result there appears an additional factor of 2 since we identify the IR and the UC cutoff via $\epsilon = 1/L$.}

\section{Supersymmetric interfaces}
\label{sec:SUSY}

Let us now consider a situation with $N=1$ supersymmetry by adding a free boson to the theory of a free fermion. 
We can  combine our results with those of  \cite{sakai_entanglement_2008} to obtain the entanglement entropy through a supersymmetric interface. 
Compatibility of the interface with $N=1$ supersymmetry requires that the interface intertwines the supercurrents $G^1, G^2$ of the two theories 
\begin{equation}
(G_r^1 - i\eta^1_S \bar{G}_{-r}^1) I_{1,2} = I_{1,2}  (G_r^2 - i\eta^2_S \bar{G}_{-r}^2) \ ,
\end{equation}
where $\eta, \eta_S^2, \eta_s^1 = \pm 1$. The signs $\eta_S^i$ specify the preserved SUSY of the two theories and the GSO projection requires to sum over both choices for $\eta$ in the final step. As explained in \cite{bachas_worldsheet_2012} the choices of sign can be absorbed in the gluing matrix $\Lambda \in O(1,1)$ for the fermions, such that specific entries in the gluing matrix for bosons and fermions can differ by signs. Since the entanglement entropy does not depend on these choices, we simply assume that bosons and fermions are glued by the same matrix $\Lambda$ (or equivalently ${\cal O}$) that we used in the current paper, implying that the preserved SUSY is the same on the two sides of the interface. 

The implementation of the GSO projection has been discussed in section \ref{sec:explicitInterfaces} and can be taken over to the supersymmetric situation. 

The full interface operator of the supersymmetric theory can be written as a tensor product of a bosonic and a fermionic piece:
\begin{equation}
 I_{1,2}(\Lambda, k_1, k_2, \varphi) = I^\text{bos}_{1,2} (\Lambda, k_1, k_2, \varphi) \otimes I^\text{ferm}_{1,2} (\Lambda) \ .
\end{equation}
Interfaces of the theory of a free boson compactified on a circle were considered in \cite{Bachas:2007td,sakai_entanglement_2008}. They depend on two integers, $k_1$ and $k_2$, specifying topological winding numbers, as well as two continuous  moduli, $\phi_1, \phi_2$. The origin of these parameters is easiest to understand in the ``folded'' picture, where the interface is mapped to a D-brane on a torus. In the simplest case of a one-dimensional brane wrapping the torus $S^1 \times S^1$, the integers can be understood as winding numbers of the D1-brane around the two $1$-cycles. The continuous moduli correspond in this picture to position and Wilson line. The gluing matrix $\Lambda$ is restricted by the torus geometry
$$
\tanh \gamma = \frac{(k_1 R_1)^2 - (k_2 R_2)^2}{(k_1 R_1)^2 + (k_2 R_2)^2} .
$$
We choose the fermionic gluing matrix to be  the same as the bosonic one. The entanglement entropy has been computed in \cite{sakai_entanglement_2008} with the result
\begin{equation}\label{eq:bosonentropy}
S^{\text{bos}}= \sigma(s)^{\text{bos}} \log L - \log |k_1 k_2| \ ,
\end{equation}
where
\begin{equation}
\sigma(s)^{\text{bos}} = \frac{s}{2} - \frac{2}{\pi^2} \int_0^\infty u \left(\sqrt{1+\frac{s^2}{\sinh^2(u)}}-1\right) du\,.
\end{equation}
The Hilbert space of the supersymmetric theory is the tensor product of the bosonic and fermionic Hilbert spaces. The partition functions $Z(K)$ hence takes the product form $Z(K) = Z^\text{bos}(K)\cdot Z^\text{ferm}(K)$. Due to the logarithm in \eqref{eq:masterformula} the entanglement entropy can then be written as
\begin{equation}
 S^\text{SUSY} = S^\text{bos} + S^\text{ferm}\,,
\end{equation}
where in the fully supersymmetric model $S^\text{bos}$ is given in (\ref{eq:bosonentropy}) and $S^\text{ferm}$ is $S^\text{NS} = S^\text{R}$.\footnote{One can also consider the GSO projection of the supersymmetric model. It has the same structure but with $S^\text{ferm}$ given by $S^\text{n.}$  or $S^\pm$. In the final result, there appears an additional contribution log\,2 for the neutral interfaces, as discussed before .} Explicitly,
\begin{equation}
S^\text{SUSY}=\frac{s}{2}  \log L - \log |k_1 k_2| \, .
\end{equation}
The prefactor $\sigma(s)$ of the logarithmic term simplifies significantly. Here, contributions from the oscillators of the bosonic and fermionic part of the system cancel out in the limit $\delta \to 0$, such that only the term $s/2$ remains. This is similar to the computations in \cite{bachas_worldsheet_2012}, where the limit of two parallel interfaces approaching each other was considered.
Note that the constant contribution $\log |k_1 k_2|$  has a topological interpretation: The winding and momentum modes of the compactified boson are quantized and part of a lattice. The combination $|k_1 k_2|$ is the index $\textit{ind}\, \hat\Lambda$ of the sublattice of windings and momenta to which the interface couples. Here, $\hat\Lambda \in O(d,d|{\mathbb Q}) $ is the gluing matrix for integer charge vectors and, as opposed to $\Lambda$, does not depend on the moduli.  On the other hand, the quantity $s = \vert\sin 2 \phi \vert = \sqrt{\mathcal{T}}$ specifies the precise geometry of the sublattice and determines the transmissivity, which is the same for bosons and fermions (for equal $\Lambda$) and also in the supersymmetric system. The supersymmetric entanglement can thus be rewritten as
\begin{equation}
S^\text{SUSY}=\frac{\sqrt{\mc{T}} c}{3}  \log L - \log \, \textit{ind}\, \hat\Lambda \ .
\end{equation}
It is very suggestive that this form of the entanglement entropy generalizes to supersymmetric torus compactifications in higher dimensions.  As was shown in \cite{bachas_worldsheet_2012}  the index of the  sublattice is a useful quantity to characterize topological information of an interface, in other words, the information that does not change under deformations of the interface or bulk theories. In a similar way, ${\cal T}$ naturally exists for any interface and characterizes the transmissivity, in other words, how far away the interface is from being  topological. In the case of higher dimensional tori, $\Lambda \in O(d,d)$ is a $2d \times 2d$ matrix consisting of $d\times d$ blocks and the transmissivity is given in terms of the determinant of the lower-right block $\Lambda_{22}$, ${\cal T} = |\det \Lambda_{22}|^{-2}$. 

\section{Conclusions and Outlook}
\label{sec:conclusions}

In this paper we have discussed the entanglement entropy through conformal interfaces for the Ising model -- i.e. a free fermion theory -- and for supersymmetric systems. We have computed the prefactor $\sigma(\mathcal{T})$ in equation (\ref{eq:result}) and seen explicitly how it arises purely from the contribution of higher oscillator modes. These largely cancel against bosonic modes in the model with supersymmetry.

It would be very interesting to generalize these findings further. It is suggestive that also in more general systems topological data of the defect will enter the constant shift $C$ in (\ref{eq:result}), whereas oscillator data enters the prefactor of the logarithmic term. 

The defects we investigated in this paper can be regarded as marginal perturbations of the topological defects of the Ising model. The latter are labelled by the primaries of the theory \cite{Petkova:2000ip}, in the case at hand  $1, \sigma, \epsilon$. The perturbing operator is a marginal defect perturbation, living only on the defect. It would be interesting to consider more generally the entanglement entropy for initially topological defects perturbed by marginal and possibly also relevant defect operators. 

Another form of perturbation appears for the interfaces in the free boson theory considered in \cite{sakai_entanglement_2008} and in section \ref{sec:SUSY} above. They come in several classes characterized by topological data $k_1, k_2$, which cannot be changed under perturbations. However, for $k_1, k_2$ fixed, there are again interfaces related by perturbations, but this time marginal bulk perturbations deforming the CFT at one side of the defect.\footnote{There are also marginal defect perturbations for the free boson, which however change neither transmissivity nor entanglement entropy.} It would be very interesting to generalize this to other systems related by RG domain walls \cite{Brunner:2007ur,Gaiotto:2012np,Konechny:2014opa,Poghosyan:2014jia} (where the free boson ``RG domain walls'' are obtained for $k_1=k_2=1$ in the above discussion) and to compute the entanglement entropy for them, e.g. in perturbation theory, similar to the discussion of the defect entropy in \cite{Konechny:2014opa}.

Apart from being interesting from the point of view of the physics of impurities, this program might also be interesting from the point of view of the physics of defects. In the discussion of defects, well-known quantites are the $g$-factor and the reflectivity/transmissivity. The entanglement entropy might provide another useful characteristic of an interface, where the transmissivity enters the prefactor of the logarithmic part whereas topological features enter separately, namely in the constant part. This is different for the $g$-factor, where both topological data and oscillator data  where they calculate the entanglement entropy up to a factor of two that appears since we identify the UV and IR cutoffenter in one factor.

\newpage
\appendix

\section{Special functions}

In the following we use $q = e^{2\pi i \tau}$.

\begin{boldmath}
\subsection{The Dedekind \texorpdfstring{$\eta$}{eta}-function}
\end{boldmath}
\noindent
The Dedekind $\eta$-function is defined as

\begin{equation}
 \eta(\tau) = q^{\frac{1}{24}} \prod_{n=1}^\infty \left(1-q^n\right)\,. \label{eq:eta}
\end{equation}

It behaves under $T$- and $S$-transformations as

\begin{equation}
 \eta(\tau+1) = e^{\frac{\pi i}{12}}\, \eta(\tau) \,,~~~~ \eta\left(-\frac{1}{\tau}\right) = \sqrt{-i\tau}\, \eta(\tau)\,.
\end{equation}

\begin{boldmath}
\subsection{The \texorpdfstring{$\theta$}{theta}-functions}
\end{boldmath}
\noindent
A most general form of the $\theta$-functions is given by

\begin{equation}
 \begin{split}
\theta \left[\alpha,\beta\right]\hspace{-2pt}(\tau,z) &= \sum_{n\in \Z} q^{\frac{1}{2}(n+\alpha)^2}\, e^{2\pi i (n+\alpha)(z+\beta)} \\
 &= \eta(\tau) \,e^{2 \pi i \alpha (z+\beta)}\,q^{\frac{\alpha^2}{2}-\frac{1}{24}}\prod _{n=1}^\infty \left(1+q^{n+\alpha-\frac{1}{2}}\,e^{2\pi i (z+\beta)}\right) \left(1+q^{n-\alpha-\frac{1}{2}}\,e^{-2\pi i(z+\beta)}\right)\,.
 \end{split}\label{eq:genTheta}
\end{equation}

Its modular transformations are given by

\begin{equation}
\begin{split}
\theta[\alpha,\beta](\tau+1,z) &= e^{-i\pi\alpha(\alpha-1)}\,\theta[\alpha,\alpha+\beta-\frac{1}{2}](\tau,z)\,,\\
\theta[\alpha,\beta](-\frac{1}{\tau},\frac{z}{\tau}) &= \sqrt{-i\tau}\,e^{2\pi i\alpha\beta+i\pi \frac{z^2}{\tau}}\,\theta[\beta,-\alpha](\tau,z)\,.
\end{split}
\end{equation}

\subsection{Bernoulli polynomials and numbers}
\label{sec:Bernoulli}

The Bernoulli polynomials $B_n(x)$ are defined by 

\begin{equation}
 \frac{t\, e^{xt}}{e^t-1} = \sum_{n=0}^\infty B_n(x)\frac{t^n}{n!}~~~~\text{with}~~\vert t\vert < 2\pi\,.
\end{equation}

The Bernoulli numbers $B_n$ are given by the polynomials evaluated at $x=0$, namely $B_n = B_n(0)$. Odd Bernoulli numbers vanish. 

The sums of $m$th powers of integers can be expressed by the use of Bernoulli polynomials and numbers as

\begin{equation}
 \sum_{k=1}^N k^m = \frac{B_{m+1}(N+1)-B_{m+1}}{m+1}
\end{equation}

Using the facts that $B_n'(x) = n B_{n+1}(x)$ and $B_n(1) = B_n$ for $n\neq 1$ and $B_1(1) = 1/2$ together with the integral representation 

\begin{equation}
 B_{2n}  = 4 n (-1)^n \int_0^\infty \frac{t^{2n-1}}{1-e^{2\pi t}}dt
\end{equation}
one can conclude that

\begin{equation}
 \frac{1}{n+1}\partial_x B_{n+1}(x)\vert_{x\rightarrow 1} = \delta_{n,0}+\frac{1}{2}\delta_{n,1} + \left(i^n-(-i)^n\right) \int_0^\infty \frac{n \,t^{n-1}}{1-e^{2\pi t}}dt\,. \label{eq:BernoulliDerivative}
\end{equation}

\section{The partition function for odd \textit{K} is enough}

\label{sec:ArgJustOdd}

We here want to show that it suffices to consider the formula for $Z(K)$ as given in \eqref{eq:oddZ} for odd $K$ to derive the entanglement entropy. We assume that the natural analytic continuation of $Z(K)$ as given in \eqref{eq:R3+}  

\begin{equation}
 Z(K) \stackrel{(a)}{=} \prod_{n>0} (p^+)^K + (p^-)^K + 2\left(\sin(2\phi) e^{-n \delta}\right)^{2K} \equiv \prod_{n>0} F_n(K)\,,
\end{equation}
gives the right result for the entanglement entropy.\footnote{This really is an assumption since the continuation is not unique.} For odd $K$ the partition function is equivalently given by \eqref{eq:oddZ} whereas for even $K$ we have to add $4 \left(\sin(2\phi) e^{-n \delta}\right)^{2K}$ to every factor. Thus the analytic continuation of the partition function has the form

\begin{align}
 Z(K) & \stackrel{(b)}{=} \prod_{n>0}\left[ I(K)\, 4 \left(\sin(2\phi) e^{-n \delta}\right)^{2K} + \prod_{k=1}^K 2 e^{-2 n \delta } \left(2 \cos^2(\nu_k) -1 + \cosh(2 n \delta )  \right) \right]\nonumber\\
  &\equiv \prod_{n>0}  H_n(K) + G_n(K)   \,,\label{eq:evenAndOdd}
\end{align}
where $I(K)$ is an analytic function interpolating between the values for odd and even $K$ with 

\begin{equation}
 I(K) = \left\{
 \begin{matrix}
                0 \text{~~~for odd~~}K\\
                1 \text{~~~for even~}K
 \end{matrix}\right.\,.
\end{equation}

We do not know the explicit form of $I(K)$. However, we can show that it suffices to consider $G_n(K)$ in \eqref{eq:evenAndOdd} and that $I'(1)$ must vanish. Let us therefore derive the EE with the help of \eqref{eq:masterformula1} for (a) and (b):

\begin{align}
 S &= \left.\left(1-\partial_K\right) \log Z(K)\right\vert_{K\rightarrow1}\nonumber\\
 &\stackrel{(a)}{=} \sum_{n>0} \left(\log F_n(1) -\frac{1}{F_n(1)}F_n'(1)\right)\\
 &\stackrel{(b)}{=} \sum_{n>0}\left( \log(G_n(1)+H_n(1))-\frac{1}{G_n(1)+H_n(1)} (G_n'(1)+H_n'(1))\right)\nonumber\\
 &= \sum_{n>0} \left(\log F_n(1) -\frac{1}{F_n(1)}(H_n'(1)+G_n'(1))\right)\,.\label{eq:laststep}
\end{align}

In the last step we use that $H_n(1) = 0$ and $F_n(1) = G_n(1)$. In the following we show that also $G'_n(1) = F'_n(1)$. Then it really suffices to solely consider $G_n(K)$ in \eqref{eq:evenAndOdd} and, thus, we can derive the EE with the formula for odd $K$. In addition we then have to require that $H'(1)= I'(1) \,4 \left(\sin(2\phi) e^{-n \delta}\right)^{2}  = 0$ which also implies that $I'(1)$ has to vanish.    

To do so we start with rewriting

\begin{equation}
\begin{split}
 G_n(K) &= \prod_{k=1}^K 2 e^{-2n\delta} \left(1+\cosh(2n\delta) - \sin^2\left(\pi\frac{k}{K}\right)\sin^2(2\phi)\right) \\
 &\equiv \exp\left(\sum_{k=1}^K g_n(k/K)\right)\,,
\end{split} 
\end{equation}
where $g_n(x) \equiv  \log\left[2 e^{-2n\delta} \left(1+\cosh(2n\delta) - \sin^2\left(\pi x\right)\sin^2(2\phi)\right)\right]$ can be expanded around $x=k/K = 0$, so that

\begin{equation}
 \begin{split}
  G_n(K) &= \exp\left(\sum_{m\ge0} \frac{g_{nm}}{K^m} \sum_{k=1}^K k^m\right)\\
  &= \exp\left(\sum_{m\ge0} \frac{g_{nm}}{K^m} \frac{B_{m+1}(K+1)-B_{m+1}}{m+1}\right)\,.
 \end{split}
\end{equation}

$B_{N}(u)$ and $B_{N}$ are the Bernoulli polynomials and numbers, respectively. Proceeding similar as in \eqref{eq:dphi} and the following one can calculate 

\begin{equation}
 G_n'(K) = \exp\left(\sum_{m\ge0} \frac{g_{nm}}{K^m} \frac{B_{m+1}(K+1)-B_{m+1}}{m+1}\right) \cdot \left(\sum_{m\ge0} \frac{g_{nm}}{m+1} \partial_K B_{m+1}(K)\right)\,,
\end{equation}
which can be massaged further in the limit $K\rightarrow1$ as

\begin{align*}
 G'(1) = &\,\exp\left(g_n(1)\right) \cdot \left(g_n(0) + \frac{1}{2}g_n'(0) + \int_0^\infty \frac{ig_n'(it)-ig(-it)}{1-e^{2\pi t}} dt\right)\\
 =&~ 2 e^{-2n\delta} \left(1+\cosh(2n\delta)\right)\bigg(\log\left[2 e^{-2n\delta} \left(1+\cosh(2n\delta)\right)\right] -\\
 &\underbrace{-\int_0^\infty \frac{2\pi (\coth(\pi t)-1)\sin(2\phi)\sinh(2\pi t)}{1+\cosh(2n\delta)+2\sin^2(2\phi)\sinh^2(\pi t)} dt}_{= \sqrt 2 \,\text{arctanh}\left(\frac{\sqrt{\cos(4\phi)+\cosh(2n\delta)}}{\sqrt2 \cosh(n\delta)}\right)\frac{\sqrt{\cos(4\phi)+\cosh(2n\delta)}}{\cosh(n\delta)} + \log(\sin^2(2\phi))-\log(4\cosh^2(n\delta))}\,\bigg)\,.
\end{align*}

At this point one only needs tedious algebraic deformations to show that the latter equals 

\begin{equation}
 F_n'(1) = e^{-2n\delta} \left(\log(e^{-2n\delta}p^+) \, p^+ + \log(e^{-2n\delta}p^-)\, p^- + 2 \log(e^{-2n\delta} \sin^2(2\phi))\,  \sin^2(2\phi)\right)\,,
\end{equation}
with $p^\pm$ given as in section \ref{sec:explNS}.


\acknowledgments

We like to thank Cornelius Schmidt-Colinet and Sebastian Konopka for helpful discussions on the topic. This work was partially supported by the Excellence Cluster Universe. 

\providecommand{\href}[2]{#2}\begingroup\raggedright\endgroup

\end{document}